# The Effect of Isotropic Pressure on the Electronic Structure and Superatomic Orbitals of Molecular $[Ag_{44}(SPhCOOH)_{30}]^{4-}$, $[Ag_{44}(SPhF_2)_{30}]^{4-}$ & $[Ag_{25}(SPhMe_2)_{18}]^-$ Nanoclusters.

**Authors:** *Lui R. Terry[†], Christopher R. Pudney[§], Henkjan Gersen[‡], Simon R. Hall[†\*].*

**Affiliations:**

[†]Complex Functional Materials Group, School of Chemistry, Cantock's Close, University of Bristol, Bristol, BS8 1TS, United Kingdom.

[§]Department of Biology and Biochemistry, Faculty of Science, University of Bath, Bath BA2 7AY, United Kingdom.

[‡]Nanophotonics and Nanophysics group, H.H. Wills Physics Laboratory, University of Bristol, Bristol, BS8 1TL, United Kingdom.

\*Correspondence to: simon.hall@bristol.ac.uk

We present the first experimental investigation of the effect of increased isotropic pressure on the superatomic electronic structure of metal nanoclusters in the molecular state. Broad multiband absorbing $[Ag_{44}(SPhCOOH)_{30}]^{4-}$, $[Ag_{44}(SPhF_2)_{30}]^{4-}$ & $[Ag_{25}(SPhMe_2)_{18}]^-$ nanoclusters were optically examined up to 200 MPa, revealing a reversible change to the superatomic electronic structure. Deviations from the ambient spectra became significant above 50 MPa, revealing both red and blue shifts to spectral features. Comparison of the spectral peaks to calculated electronic transitions indicate that electronic states on the ligand are destabilized and that the splitting of the superatomic orbitals decreases with increasing pressure. These findings highlight that under relatively modest pressure the fundamental superatomic electronic structure of metal nanoclusters can be manipulated, thus the optical and electronic properties of functional materials containing nanoclusters such as in solar cells and photocatalytic devices can be tuned by the application of pressure.

# 1. Introduction

Metal nanoclusters are a class of nanomaterial that is composed of less than a few hundred atoms and can be considered molecular-like. Typically < 3 nm, they are described by the nuclearity of their core (number of atoms) instead of their core diameter *i.e.* $Au_{25}$, $Zn_{256}$, $Al_{13}$, $Ag_{155}$ etc. Sizes larger than this are termed metal nanoparticles and typically can range from 3-100+ nm, hence are often described in the literature by their physical diameters rather than nuclearity. In recent years, synthetic efforts have given precise control over the composition and size of metal nanoclusters, producing a library of chemically stable metal nanoclusters containing a protective organic-ligand surface layer, such as $[Ag_{44}(SPhCOOH)_{30}]^{4-}$,[1–4] $[Ag_{44}(SPhF_2)_{30}]^{4-}$,[5,6] $[Ag_{25}(SPhMe_2)_{18}]^{-}$,[7] $[Au_{25}(SR)_{18}]^{-}$ [8–10] & $Au_{102}(SR)_{44}$ [11–14] allowing for facile manipulation and analysis in the laboratory. Experimental study of protected metal nanoclusters has revealed interesting optical properties such as two photon absorption[15,16] and photo emission[17] as well as magnetic properties[18] and are emergent materials in energy, environmental and health based applications[19] such as bioimaging[20], chemical sensing[21] and photovoltaic devices[22]. The molecular-like state of these nanoclusters opens up the possibility to form emergent functional materials based on the nanocluster building blocks[23] as recently demonstrated by *Yang et al,*[5] where the broad multiband absorbing $[Ag_{44}(SPhF_2)_{30}]^{4-}$ nanoclusters were crystallized producing a stable nanocluster-based material $[4PPh_4][Ag_{44}(SPhF_2)_{30}]·(6CH_2Cl_2)$. Furthermore, many of the ligand protected metal nanoclusters have been crystalized into a nanocluster-based material with known crystal packing structures.[24] The solid-state crystals of this type of novel nanocluster-based material may be useful in future solar cell technologies[22,25–28] or molecular electronics,[29–31] the physical properties of which depend on the electronic structure and properties of the individual nanoclusters.

The fulcrum of intrigue for both protected and unprotected metal nanoclusters is that an electronic shell structure is known to exist,[32–34] where the valence electrons of the metal core occupy a set of "superatomic" orbitals. The low level of nuclearity and therefore low electron valency in metal nanoclusters, affects the density of states of the nanocluster such that; the quasi continuous energy levels that are present in larger metallic nanoparticles and the bulk state now becomes discrete and molecular-like. The collective electron oscillation (Plasmon resonance) observed in metal nanoparticles disappears and a distinctive optical absorption spectrum is observed. Superatomic orbitals are analogous to atomic orbitals, where the same spherical harmonics (S,P,D,F… *etc*.) are exhibited and a closed shell confers stability, but are defined by the entire nanocluster of atoms instead of individual atoms. Consequently, superatomic orbitals are delocalized across the spherical core of the nanocluster, therefore, the radial extension of superatomic orbitals is relatively large compared to their atomic equivalents. Electron filling of these delocalized states closely follow a partial Jellium model[35] adhering to the Aufbau principle, where shell closings follow the format: $1S^2|1P^6|1D^{10}|2S^2|1F^{14}|...$ etc. Thus, variation in nanocluster size (or electron valency) results in different superatomic orbitals being filled, theoretically giving rise to different electronic, magnetic, optical and oxidative properties of the individual nanoclusters.

To date little is known about the collective optical and electronic properties of novel metal nanocluster-based materials. Both optical and electronic properties of the nanoclusters may be altered due to proximity effects as has been seen previously for $C_{60}$ molecules via low temperature scanning tunneling microscopy experiments,[36–38] where observed superatomic states delocalized across a framework of $C_{60}$ molecules with nearly free electron like character. Pressure experiments by *Li et al* [39,40] on arrays of colloidal plasmonic sized metallic nanoparticles (5.5 nm) show a red-shift of the plasmon resonance with increasing pressure, indicating collective behavior between the nanoparticles with decreasing separation. Whether a similar collective behavior occurs for nanoclusters is in question. Pressure based experiments provide a useful platform for screening how the intermolecular separation of these nanoclusters affects the electronic and optical properties of the collective solid without changing the chemical composition of the nanocluster itself. The properties of such metal nanocluster-based material will depend on the properties of the individual building blocks in the same way that the properties of common materials depend on the properties of their constituent atoms. As a first step, it is therefore crucial to obtain a better understanding of the optical and electronic properties of the individual metal nanocluster and how it is affected by pressure before examining the solid state. The effect of pressure on the electronic structure of superatomic materials has been studied on solid state systems such as Fullerene single crystals,[41,42] but has not been studied previously on the molecular state of superatomic metal nanoclusters. To the best of our knowledge, pressure-based studies of metal nanoclusters have so far been restricted to simulations only.[3]

In this work, we report the first experimental optical study of molecular metal nanoclusters under pressure, more specifically we show how the application of relatively modest isotropic pressure affects the UV-Vis absorption spectrum and therefore electronic structure of spherical $[Ag_{44}(SPhCOOH)_{30}]^{4-}$, $[Ag_{44}(SPhF_2)_{30}]^{4-}$ & $[Ag_{25}(SPhMe_2)_{18}]^-$ nanoclusters in their molecular state. The three nanoclusters chosen each exhibit a broad multiband absorption spectrum and unique electronic structure which could be utilized in future solar cell technology or molecular electronics. Each metal nanocluster chosen exhibits electronic shell structure originating from the overlap of the valence 5s electrons provided by the silver atoms. Nanoclusters with the same valence electron count (and therefore same filled superatomic orbitals) have been chosen but with differing surface ligands ($[Ag_{44}(SPhCOOH)_{30}]^{4-}$ and $[Ag_{44}(SPhF_2)_{30}]^{4-}$), as well as a nanocluster with a lower valence electron count ($[Ag_{25}(SPhMe_2)_{18}]^-$) to consider how the differing superatomic orbitals and ligands contributing to the electronic structure are affected by external pressure. Nanoclusters were also chosen for their relative atmospheric stability and facile synthesis. Our results indicate that increased pressure favors the destabilization of the occupied ligand orbitals and decreases the energy splitting of superatomic orbitals for all nanoclusters examined. This information informs us that the fundamental superatomic energy levels of metal nanoclusters can be manipulated on a molecular basis under fairly modest conditions, thus the optical/electronic properties and performance of a functional material containing nanoclusters can be tailored by the application of modest pressure.

## 2. Results

The chosen metal nanoclusters $[Ag_{44}(SPhCOOH)_{30}]^{4-}$, $[Ag_{44}(SPhF_2)_{30}]^{4-}$ & $[Ag_{25}(SPhMe_2)_{18}]^-$ were dissolved in dimethylformamide (DMF) at low concentration and examined by UV-Vis spectroscopy at 15 °C up to a maximum isotropic pressure of 200 MPa, followed by a return to ambient pressure. The only other alternative solvent for these three nanoclusters, dimethyl sulfoxide (DMSO), was also attempted across this range to expand the investigation, however it solidified under these conditions preventing further experimentation. Increasing isotropic pressure to 200 MPa resulted in an observable change in the molecular absorbance spectrum of the nanoclusters (**Figures 1,2 & 3**) with deviations from the ambient spectra, becoming significant above 50 MPa. Indicating that the molecular electronic energy levels of the nanoclusters shifted with the changing pressure at remarkably modest pressures. Spectral change with respect to pressure is emphasised in the difference spectrum shaded grey (absorbance difference between maximum and ambient pressure). For all clusters, the trends in spectral shift were found to be essentially reversible on decreasing the pressure to atmospheric, with no new spectral features observed, indicating there is no permanent alteration of the nanoclusters at this temperature and pressure range. Changes in the exact peak position with pressure can be found in the supporting information (SI) Figures S1-3. Furthermore, upon return to atmospheric pressure, both $[Ag_{44}(SPhCOOH)_{30}]^{4-}$ & $[Ag_{25}(SPhMe_2)_{18}]^-$ spectra returned to their initial level of absorbance, indicating that no insoluble aggregates were formed due to the pressure increase. However, a decrease in

**Table 1.** Optical absorption transition data of nanoclusters and the associated transition energy change on increasing isotropic pressure to 200 MPa.

| Nanocluster | Observed peak positions (nm) | Calculated peak positions‡ (nm) | Transition from occupied orbital‡ | Transition to unoccupied orbital‡ | Observed peak shift |
|---|---|---|---|---|---|
| [Ag$_{44}$(SPhCOOH)$_{44}$]$^{4-}$ | 830 | 837 | HOMO (1D); Ag/ligand states | LUMO+1 (1F); LUMO (2S) | undetected |
| | 646 | 670 | Ligand states; Oxygen | LUMO+1 (1F); LUMO (2S) | Red |
| | 542 | 553 | Oxygen/carbon; Oxygen | LUMO+1 (1F); LUMO (2S) | Red |
| | 490 | 476 | Carbon | LUMO+1 (1F) | none |
| | 422 | 444 | Carbon; HOMO (1D) | 2S & 1F; Carbon/Oxygen | Red |
| [Ag$_{44}$(SPhF$_2$)$_{30}$]$^{4-}$ | 833 | 954 | HOMO (1D) | LUMO+1 (1F) | undetected |
| | 638.5 | 642 | Ligand states; HOMO (1D) | LUMO+1 (1F); LUMO+1 (1F) | Red |
| | 532.5 | 561 | - | - | Red |
| | 480.5 | 500 | HOMO (1D); ligand; | Ligand states; LUMO+1 (1F); | Blue |
| | 412.5 | 416 | HOMO (1D); ligand; Ag (4d) | Ligand states; LUMO+1 (1F); Ag (5sp) | none |
| [Ag$_{25}$(SPhMe$_2$)$_{18}$]$^{-}$ | 686 | 756 | HOMO (1P) | LUMO (1D) | Blue |
| | - | 532 | HOMO (1P) | LUMO+1 (1D) | - |
| | 494.5 | 498 | HOMO-1 (ligand/sulphur) | LUMO (1D) | Red |
| | - | 468 | HOMO (1P) | LUMO+3 (Mixed ligands) | - |
| | 392.5 | 395 | HOMO-1 (ligand/mixed) | LUMO+1 (1D) | Red |
| | 335 | - | - | - | Blue |

a) Calculated peak positions and associated transitions are taken from references [2][5,6][43-46]

absorbance relative to the initial spectra was observed upon return to atmospheric pressure for [Ag$_{44}$(SPhF$_2$)$_{30}$]$^{4-}$ nanoclusters, potentially indicating the formation of insoluble aggregates upon increased pressure for this nanocluster. The cycled pressure spectra can be found in the SI Figures S4-6.

An insight into the observed spectral shifts as a function of pressure was gained by cross examining with the electronic structure of the nanoclusters given by time-dependent density functional perturbation theory (TD-DFPT) calculations performed by *Aikens*,[43,44],[45] *Yoon & Landman*,[2] *Gell & Hakkinen*[5,6] *and Tlahuice-Flores*.[46] Several of the spectral features have been previously correlated to optical transitions by TD-DFPT, the associated calculated transitions are highlighted in **Table 1**. To elucidate how the nanoclusters electronic energy levels are affected by increased pressure, the calculated electronic transitions are expounded in a Kohn-Sham diagram from the calculated projected density of states and aligned to the associated wavelengths of the optical spectra above in Figures 1, 2 & 3. Thus, any shifts to the observed peaks can be easily cross-examined against the associated calculated transitions. The electronic transition arrows have been coloured to represent the observed experimental shifting with pressure of the associated spectral features (*i.e.*: red = red shift, blue = blue shift, black = no observed shift). Firstly, two nanoclusters identical in nuclearity and electron valence count but with differing ligand systems were investigated, $[Ag_{44}(SPhCOOH)_{30}]^{4-}$ & $[Ag_{44}(SPhF_2)_{30}]^{4-}$. Changes to their spectra (Figures 1 and 2 respectively) with increased isotropic pressure are herein discussed.

$[Ag_{44}(SPhCOOH)_{30}]^{4-}$ is a spherical Keplerate solid composed of a hollow icosahedral $Ag_{12}$ core enclosed by an $Ag_{20}$ dodecahedron, octahedrally capped with $Ag_2(SR_5)$ ligand mounts[1]. Containing 18 valence electrons occupying superatomic orbitals in the electron configuration $1S^2|1P^6|1D^{10}|$, where its highest occupied molecular orbital (HOMO) is the $1D^{10}$ superatomic orbital split into a lower energy doubly degenerate set of orbitals and a triply degenerate set of orbitals of higher energy. The lowest unoccupied molecular orbital (LUMO) is the $2S^2$ superatomic orbital with a calculated HOMO-LUMO energy gap ($\Delta_{HL}$) = 0.78eV.[1] Its full electronic energy diagram is illustrated in Figure 1. The $[Ag_{44}(SPhF_2)_{30}]^{4-}$ nanocluster has a similar crystal structure to $[Ag_{44}(SPhCOOH)_{30}]^{4-}$ however, the para-mercaptobenzoic acid ligands are replaced with 3,4-difluorobenzenethiol ligands. Equally, the valence electron count is also 18, occupying the same superatomic orbitals with a calculated $\Delta_{HL}$ = 0.77 eV, its full electronic energy diagram is illustrated in Figure 2. The TD-DFT calculations performed by *Gell & Hakkinen*[5,6] and *Yoon & Landman*[2] utilize the exact crystal structure of the nanoclusters used in these experiments giving accurate representations of the optical transitions and their origins. Both nanoclusters exhibit a similar optical absorption spectra observing subtle changes with increased pressure as highlighted in Figure 1 & 2. Both $Ag_{44}$ nanoclusters exhibit their first optical transitions at ~830 nm pertaining to the HOMO-LUMO+1 transition. Discerning any true shift of this transition was obscured due to the broadness of the peak and the low concentration used to visualize all peaks below an absorbance of 1, but could be observed in a future experiment at higher concentrations at the expense of other spectral peaks.

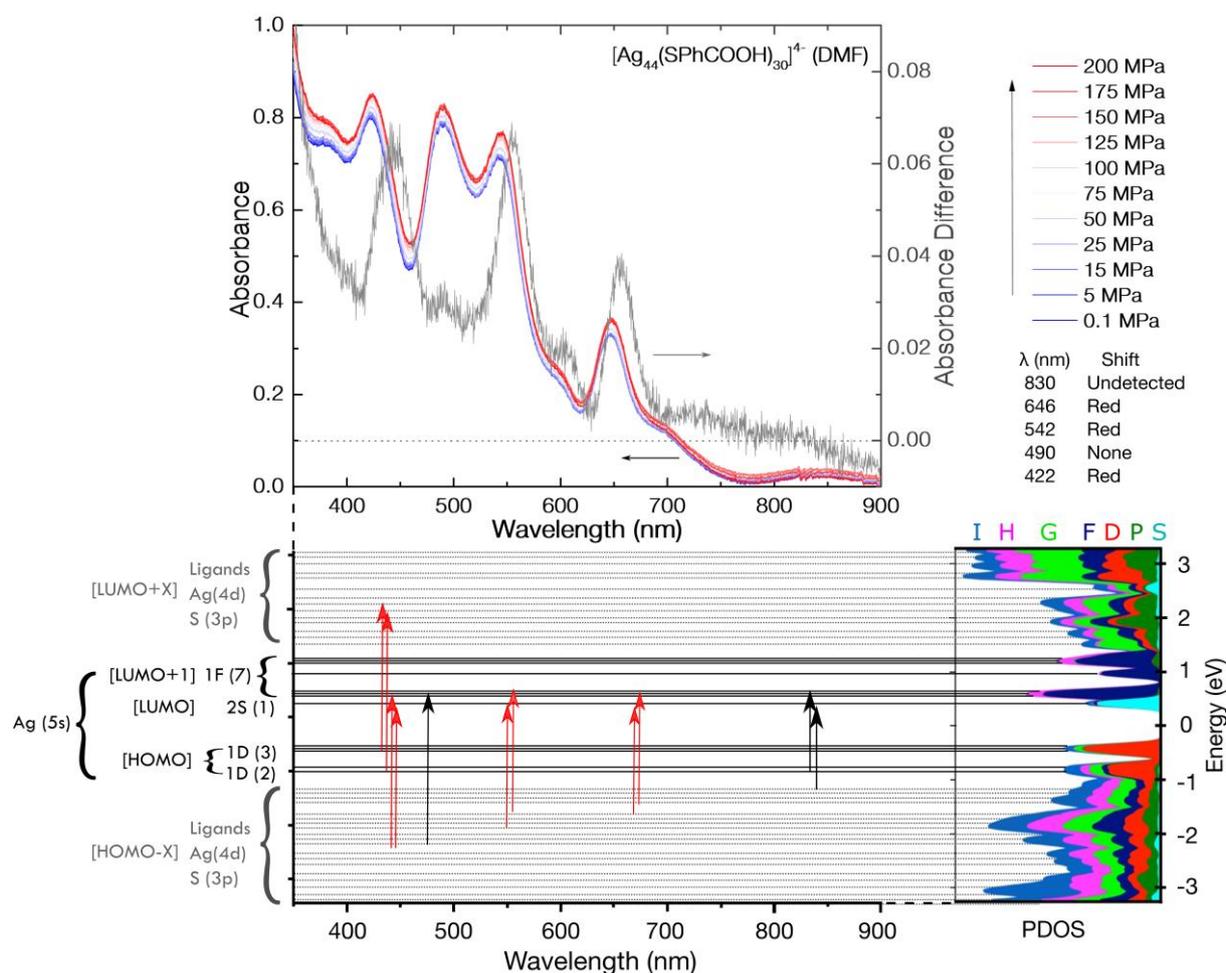

**Figure 1.** UV-Vis absorption spectra of nanoclusters [$Ag_{44}(SPhCOOH)_{30}$]$^{4-}$ in DMF with increasing isotropic pressure (blue to red). The right axes highlight the difference spectra (grey) between the ambient pressure and maximum pressure absorbance. To help visualize the change to the nanoclusters electronic energy system, the projected density of states (PDOS) and calculated energy level transitions of the nanocluster (Adapted with permission from *Conn et al*[2], Copyright (2015), American Chemical Society) are abstracted into a Kohn-Sham diagram, where calculated electronic transitions are aligned to their associated wavelengths in the graph above.

For both $Ag_{44}$ nanoclusters, the second and third spectral features (~640 & 530 nm) show a decrease in transition energy with increasing pressure, however, the calculated origins of the electronic transitions differ between the two species. For $[Ag_{44}(SPhF_2)_{30}]^{4-}$ nanoclusters the second spectral feature is reduced in energy (red shifted) and mainly consists of a transition from occupied states on the ligands to the superatomic 1F with some core-core transition character from the occupied superatomic 1D to the superatomic 1F orbitals. In comparison, the $[Ag_{44}(SPhCOOH)_{30}]^{4-}$ nanoclusters second spectral feature is red shifted and contains no core-core transitions. It is composed of a transition from occupied states on the ligands to the superatomic 1F orbital and a transition from occupied states on the ligands to the superatomic 2S orbital. The decrease in energy of this optical transition suggests occupied states on the ligands are destabilised (move to higher energy) and/or the unoccupied states on the 1F superatomic orbital are stabilised (move to lower energy). The third spectral feature in $[Ag_{44}(SPhF_2)_{30}]^{4-}$ nanoclusters is composed of a transition from lower energy states on the ligands to the superatomic 1F orbital and is red shifted. A similar transition is observed for the $[Ag_{44}(SPhCOOH)_{30}]^{4-}$ nanocluster with contributions from lower energy states on the ligands to both the superatomic 1F and 2S orbitals and is red shifted. A significant observation of the $Ag_{44}$ nanoclusters spectra with increasing pressure is their fourth spectral feature at ~480 nm. For $[Ag_{44}(SPhF_2)_{30}]^{4-}$ nanoclusters an increase in energy (blue shifted) is observed for this transition from the 1D superatomic orbital to unoccupied states on the ligand. In contrast, no change is observed for the $[Ag_{44}(SPhCOOH)_{30}]^{4-}$ nanoclusters fourth spectral feature. Its' transition at ~480 nm is similar to the previous spectral features, originating from low energy states on the ligands to the superatomic 1F orbital but differs with no contribution into the superatomic 2S orbital. The final spectral feature at ~420 nm also differs between the two nanoclusters. $[Ag_{44}(SPhF_2)_{30}]^{4-}$ observes no shift and is composed of transitions between states on the ligand to the 1F orbital (comparable to the 480 nm spectral feature of $[Ag_{44}(SPhCOOH)_{30}]^{4-}$), some transitions from the 1D orbital to unoccupied states on the ligand as well as interband transitions from Ag (4d) to Ag (sp).[6] In contrast, the $[Ag_{44}(SPhCOOH)_{30}]^{4-}$ nanoclusters transition experienced a red shift, however, no interband transition occurs, only ligand to core and core to ligand transitions.

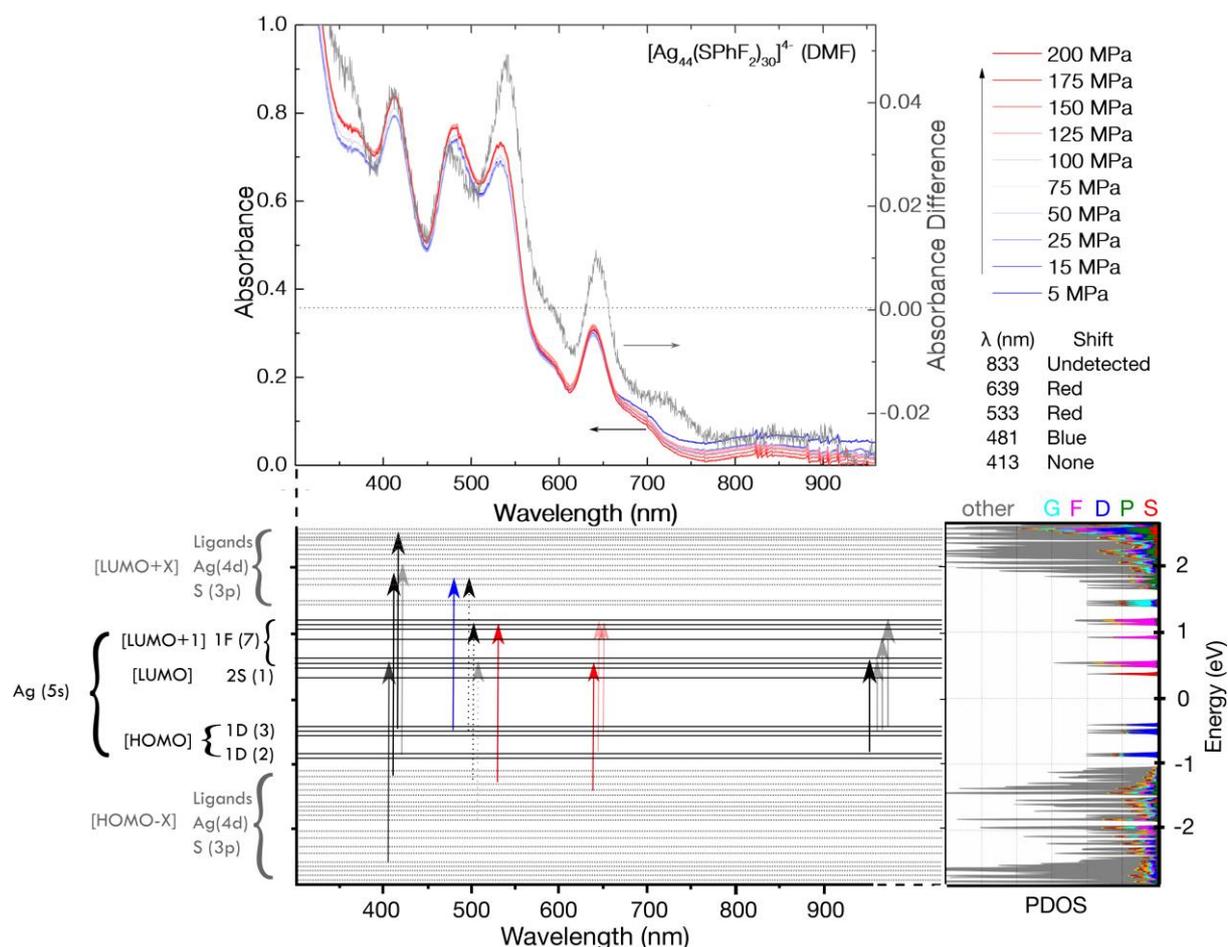

**Figure 2.** UV-Vis absorption spectra of nanoclusters $[Ag_{44}(SPhF_2)_{30}]^{4-}$ in DMF with increasing isotropic pressure (blue to red). The right axes highlight the difference spectra (grey) between the ambient pressure and maximum pressure absorbance. To help visualize the change to the nanoclusters electronic energy system, the projected density of states (PDOS) and calculated energy level transitions of the nanocluster (Reprinted by permission from *Yang et al*[5], copyright (2013), Macmillan Publishers Ltd: Nature Communications) are abstracted into a Kohn-Sham diagram, where calculated electronic transitions are aligned to their associated wavelengths in the graph above.

Next, a smaller nanocluster with a lower valence electron count was investigated to give a broader understanding of the effects of pressure. $[Ag_{25}(SPhMe_2)_{18}]^-$ is composed of a $Ag_{13}$ icosahedral core capped with six V-shaped $Ag_2(SR)_3$ ligand units in a pseudo-octahedral motif.[7] Containing a total of 8 valence electrons occupying superatomic orbitals in the electron configuration $1S^2|1P^6|$. Where the HOMO is the $1P^6$ and the LUMO is the $1D^{10}$ superatomic orbital split into a lower energy doubly degenerate set of orbitals (LUMO) and triply degenerate set of orbitals of higher energy (LUMO+1). Its calculated $\Delta_{HL}$ = 1.64 eV.[43] Unfortunately, TD-DFT calculations have not been previously conducted on the exact crystal structure of this nanocluster, thus no PDOS is reproduced in Figure 3. Previous TD-DFT calculations have been conducted on a $[Ag_{25}(SR)_{18}]^-$ structure with differing ligands,[43–46] and have been utilized for examining the changes to the spectra. Therefore, some disparity between the experimental and theoretical data can be seen in Figure 3 due to the alternative ligands and non-precise atomic co-ordinates in simulating the spectrum and its transitions. However, although the calculations are not an exact match, they can still provide a qualitative guide to observed changes.

The first allowed optical transition (~686 nm) pertaining to a transition between the HOMO and LUMO; blue shifted with increasing pressure, suggesting either a stabilization of the superatomic 1P orbital or destabilization of the superatomic 1D LUMO. The next optical transition (~495 nm) is red shifted on a similar scale. The origin of this transition is not well defined due to the differing ligands used in the calculations of $[Ag_{25}(SR)_{18}]^-$ by *Aikens*[43–45] and *Tlahuice-Flores*[46] altering the energy levels position. However, it is highly likely to be from one or a component of several of the transitions close to the observed peak. Interestingly, it can be observed that the majority of calculated transitions <650 nm involve occupied states on the sulfur/ligands. The next optical transition (~393 nm) is also redshifted. This optical transition matches well with the calculated transition between ligand and LUMO+1. The final optical transition (~335 nm) observes an increase in energy, this transition is not defined in the literature. Transitions at this high an energy often are from low energy ligand states but can also come from core orbitals donating into higher energy ligand states.

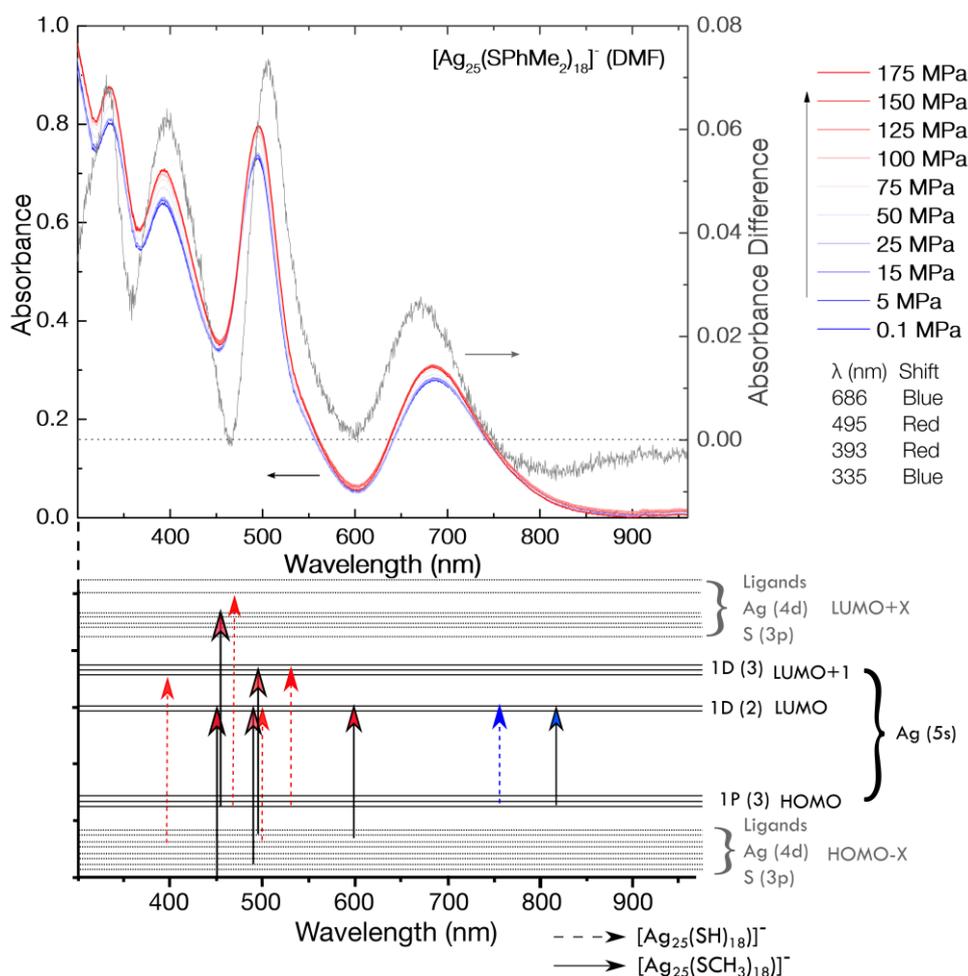

**Figure 3.** UV-Vis absorption spectra of [Ag$_{25}$(SPhMe$_2$)$_{18}$]$^-$ nanoclusters in DMF under increasing isotropic pressure (blue to red). The right axis highlights the difference spectra (grey) between the ambient pressure and maximum pressure absorbance. To help visualize the change to the nanoclusters electronic energy system, a Kohn-Sham diagram of an [Ag$_{25}$(SR)$_{18}$]$^-$ nanoclusters energy levels is provided below. Because no TD-DFT has been conducted on this exact crystal structure, no PDOS could be utilized. Thus, calculated electronic transitions are from analogous nanoclusters [Ag$_{25}$(SH)$_{18}$]$^-$ and [Ag$_{25}$(SCH$_3$)$_{18}$]$^-$ (reproduced from *Aikens*[43,45] and *Tlahuice-Fores*[46] respectively) and are aligned to the associated wavelengths in the graph above.

## 3. Discussion

The experimental results of all nanoclusters could be affected in two ways. 1) The increased isotropic pressure causes a change in the nanoclusters surrounding environment *i.e.* the solvent coordination sphere, therefore the electrostatic interaction is altered. 2) The increased isotropic pressure or force on the nanoclusters causes a physical change to its crystal structure. Both of which would influence the final electronic structure of the molecule affecting the optical absorption spectra. Analysing all the spectral shifts of the nanoclusters observed, a few qualitative observations can be drawn.

- Transitions involving occupied states from the ligands appear to be redshifted with increased isotropic pressure.
- Changes to transitions involving split states on the superatomic 1D orbital and 1F orbitals appear to differ in sign depending on which set is involved. Indicating a change in the splitting energy of the superatomic states with increased isotropic pressure.
- Deviations from the atmospheric spectra begin to appear at 25 MPa but become distinct from 50 MPa and above.
- Spectral shifts are larger for $[Ag_{44}(SPhCOOH)_{30}]^{4-}$ than $[Ag_{44}(SPhF_2)_{30}]^{4-}$.

Considering the surface of the nanocluster, the destabilization of the occupied states of the ligands is highly probable in this experiment. Because the ligands are hydrodynamic in solution, increasing the pressure may cause them to move into energetically unfavourable conformations. Consequently, this would cause the electronic states on the ligands to destabilise (move to higher energy) causing an observed red-shift to some of the peaks in the spectra, it is important to note that this would in turn stabilise (move to lower energy) unoccupied states on the ligands. Furthermore, Bader-charge analysis on many nanoclusters and those examined in this experiment suggest that the majority of anionic charge is located at the surface of the nanoclusters and not the core. In particular, most negative charge of $[Ag_{44}(SPhCOOH)_{30}]^{4-}$ is located on its benzoic acid ligands, on oxygen.[2] Similarly, the majority of negative charge is located on the fluorine atoms[5] of $[Ag_{44}(SPhF_2)_{30}]^{4-}$ ligands'. Therefore, increasing the pressure is likely to cause penetration and increased interaction of the solvent into the charge sphere of the nanocluster causing an effect similar to solvatochromism. This may also explain the difference in the magnitude of the spectral shifts between the two $Ag_{44}$ nanoclusters. An increased stabilization energy between ligand pairs due to dipole-dipole interactions[5] in the $[Ag_{44}(SPhF_2)_{30}]^{4-}$ nanoclusters may help prevent the distortion of the ligand conformation / penetration into the charge sphere and hence minimise the observed spectral change. Additionally, ligand interactions could generate nanocluster dimers or oligomers at increased pressure such as the hydrogen bonding found in solid state [$Ag_{44}(SPhCOOH)_{30}$]$^{4-}$. This could account for the increased red shift for the transitions involving the occupied states on the ligand/oxygen for this molecule, assuming the oligomers are soluble and separate upon return to ambient pressure. [$Ag_{44}(SPhF_2)_{30}$]$^{4-}$ experiences a decrease in its absorbance upon return to atmospheric pressure possibly indicating formation of insoluble aggregates in solution with the increased pressure, consequently decreasing the amount of observable shift. Considering these ideas about the surface of the nanoclusters, the ligands therefore might act as a receptor for the molecule, therefore optical analyses of the nanoclusters spectral features may indirectly give an indication of the state of ligands and their surrounding environment.

Regarding the core crystal structure, interdigitation of the solvent between the ligands and increased pressure upon the core could cause a reduction in its size *i.e.* reduced metal-metal bond lengths. A minor reduction in core radius is observed in previous high pressure quantum mechanical simulations conducted on a solid state [$Ag_{44}(SPhCOOH)_{30}$]$^{4-}$ superlattice by *Yoon et al.*[3] Most notably the ligands were shown to exhibit a large amount of flexure in the system with the increased pressure, causing them to buckle and changing the hydrodynamic radius of the nanocluster. At the lower pressures studied by *Yoon et al* a small contraction in the metal core radial distances is indeed observed (*viz.* Figure S2 reference [3]). Our work is of singular nanoclusters in the coordinating solvent which allows for the possibility for solvent molecules to interdigitate the ligands giving an increased level of pressure upon the core. Whereas, in the quantum mechanical simulation the nanocluster is locked into conformation by the extended superlattice of nanoclusters in the solid state; thus, the main pressure points are positioned where the nanoclusters link via hydrogen bonding, applying most strain on the ligands. Thus, a decrease in metal core distances of the nanoclusters is likely with the increased isotropic pressure. Theoretical studies conducted by *Aikens*[44] revealed that splitting of the superatomic 1D orbital in $Au_{25}$ and $Ag_{25}$ nanoclusters was sensitive to the metal-metal bond lengths. A decreased metal-metal bond length resulted in a decrease in the 1D splitting. This finding could help explain the observed blue shifts and zero-shifts for the nanoclusters, indicating changes in the splitting of the superatomic orbitals. For instance, In the $Ag_{25}$ nanocluster, a reduction in the splitting of the 1D orbitals could cause the observed blue shift in the first spectral feature at ~686 nm. Furthermore, in [$Ag_{44}(SPhF_2)_{30}$]$^{4-}$ all spectral features involving both sets of the HOMO superatomic 1D orbitals would not see a change if the related unoccupied orbitals remained at the same energies, but those involving the highest occupied energy set would see a blue-shift. Interestingly, all spectral features transitions in [Ag$_{44}$(SPhF$_2$)$_{30}$]$^{4-}$ from 1D use both sets except the excitation at 480 nm (*viz.* figure S9 in ref [6]), which uses only the highest energy set, accounting for the observed blue shift. Finally, altering the confirmation of the apparent octahedral crystal field of ligands around the metal core may also modify the splitting of the superatomic orbitals 1D and 1F similar to transition metal coordination complexes.

Complete analysis of the changes to the electronic structure remains challenging and additional computational experimentation is required to accurately identify the contribution of all of the outlined factors to the apparent change in the optical spectra of the nanoclusters.

However, the results provide an insight into the response of metal nanoclusters to external stimuli and that it occurs at relatively modest pressures. A diagram of the hypothesized general changes to the electronic energy levels causing the observed optical shifts with increased isotropic pressure is highlighted in **Figure 4**.

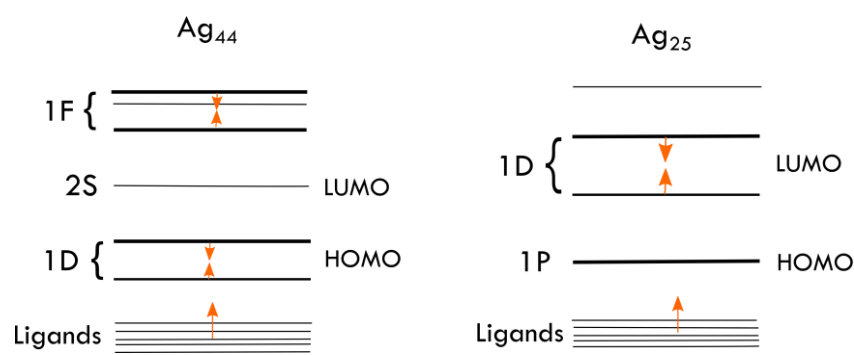

**Figure 4.** Diagram of the theorized changes to the electronic energy levels of the two nanoclusters sizes examined. The orange arrows represent the shifting of electronic energy levels to lower or higher energy with increased pressure

## 4. Conclusion & outlook

In summary, to the best of our knowledge, the first experimental investigation of the effect of increased isotropic pressure upon the electronic structure of metal nanoclusters; [Ag$_{44}$(SPhCOOH)$_{30}$]$^{4-}$, [Ag$_{44}$(SPhF$_2$)$_{30}$]$^{4-}$ & [Ag$_{25}$(SPhMe$_2$)$_{18}$]$^{-}$, in the molecular state by UV-Vis spectroscopy was conducted. An observable change to the nanoclusters spectra was discovered with increased pressure. By correlating the spectral peaks to the associated calculated electronic transitions, an understanding of the changes to the electronic energy levels was obtained. In general, the increased pressure appeared to destabilize occupied states on the ligand, as well as decrease splitting of the superatomic states for all nanoclusters observed. Several possible contributing factors responsible for the observed changes are described and hypothesized. The observed changes demonstrate that with relatively modest pressure the fundamental electronic energy levels of metal nanocluster can be manipulated. The implications of this finding suggest that the optical/electrical properties and performance of materials incorporating metal nanoclusters in their design such as solar cells, photocatalytic devices and molecular electronics could be tuned and improved by the application of modest pressure. Furthermore, this has implications towards any future pressure studies on nanocluster-based material in the solid-state and should be considered when observing changes to the electronic spectra of these superatomic materials. Further understanding of the unique electronic system found in metal nanoclusters and how it is affected by external stimuli could be made in future pressure-based experimentation by analyzing how non-spherical nanoclusters are affected by such isotropic pressure, solvent or pH dependence and determination of the HOMO position by photoemission experiments.

## 5. Experimental Section

*Synthesis & Characterisation*

The aforementioned metal nanoclusters were synthesized and then crystallised as described by *Desireddy et al,*[1] *Yang et al*[5] and *Joshi et al.*[7] The crystals were examined via single crystal X-ray diffraction (SCXRD) on a Bruker Microstar equipped with a Cu-Kα radiation source at 120 K revealing the correct unit cell dimensions for the nanocluster crystals. Nanoclusters were also examined via nanospray mass spectroscopy (MS) on a Waters Synapt G2S time-of-flight instrument with the spray nozzle set at a voltage of 1.4 kV and gas pressure of 0.4 psi. Nanospray MS revealed the parent ion peaks of the nanoclusters (Figure S7-9).

*Pressure studies*

The metal nanoclusters were prepared for UV-Vis spectroscopy measurements across the range of 0.1 - 0.2 mg ml$^{-1}$ in a solution of dry DMF. Solutions were prepared immediately prior to absorption measurements. High-pressure static absorption measurements were monitored using a Cary60 UV-Vis spectrometer (Agilent technologies) with a thermostated hydrostatic high-pressure cell (ISS High-Pressure Cell System) placed directly in the spectrometer via a custom mounting where the path length of the cell used was 1 cm and the slit width of the spectrometer was 5 nm. Each absorption spectrum was baseline corrected with dry DMF. Spectra were monitored from 350 to 900 nm at constant temperature (15 °C) and each pressure was equilibrated for 2 minutes prior to acquiring spectra. Changes to the optical spectrum were recorded up to an isotropic pressure of 200 MPa and upon return to ambient pressure. A small absorption increase is expected with increasing pressure as the local concentration of the nanoclusters in the beam will increase slightly due to a reduction in solvent volume (1.40 to 1.31 mL) calculated from the compressibility of the solvent (Figure S10).[47,48]

**Supporting Information**
Details of the calculated density change of DMF with increased pressure at 15°C and the cycled pressure spectra included in the supporting information.

**Acknowledgements**

The authors would like to thank Jason Potticary and Julia Walton for their advice on the analysis of the spectra as well as Dragana Catici and Hannah Jones for their assistance with the pressure cell. HG and CRP would like to acknowledge the contribution of the GW4 network from which this work was initiated. The work was supported in part by a grant from the USAF European Office of Aerospace Research and Development (FA8655-12-1-2078) and the EPSRC (1247228).

Supporting Information

**The effect of isotropic pressure on the electronic structure and superatomic orbitals of molecular [Ag$_{44}$(SPhCOOH)$_{30}$]$^{4-}$, [Ag$_{44}$(SPhF$_2$)$_{30}$]$^{4-}$ & [Ag$_{25}$(SPhMe$_2$)$_{18}$]$^-$ nanoclusters.**

*Lui R. Terry, Christopher R. Pudney, Henkjan Gersen, Simon R. Hall*[*].*

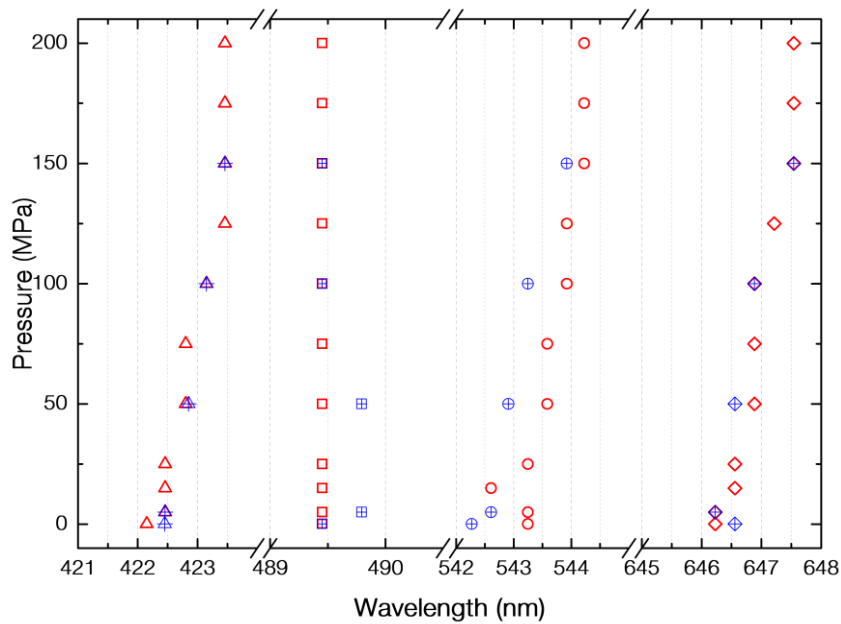

**Figure S1:** Optical absorption spectral peak centers for [Ag$_{44}$(SPhCOOH)$_{30}$]$^{4-}$ with increasing pressure in red and decreasing isotropic pressure blue.

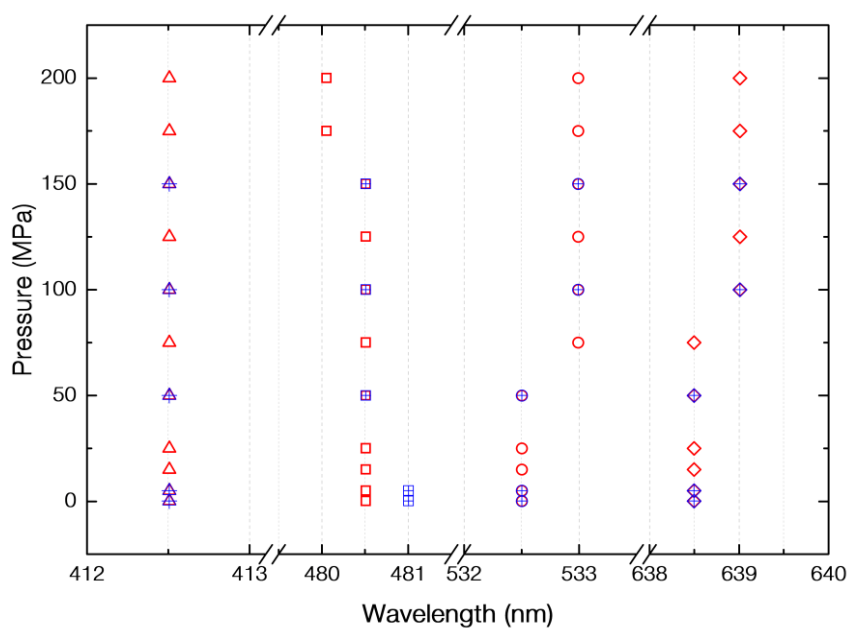

**Figure S2:** Optical absorption spectral peak centers for [Ag$_{44}$(SPhF$_2$)$_{30}$]$^{4-}$ with increasing pressure in red and decreasing isotropic pressure blue.

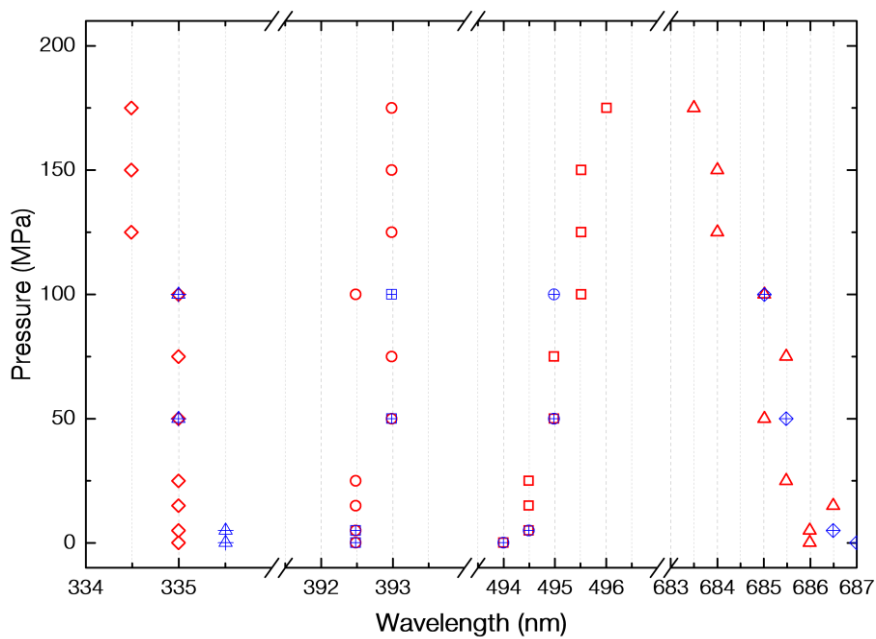

**Figure S3:** Optical absorption spectral peak centers for [Ag$_{25}$(SPhMe$_2$)$_{18}$]$^-$ with increasing pressure in red and decreasing isotropic pressure blue.

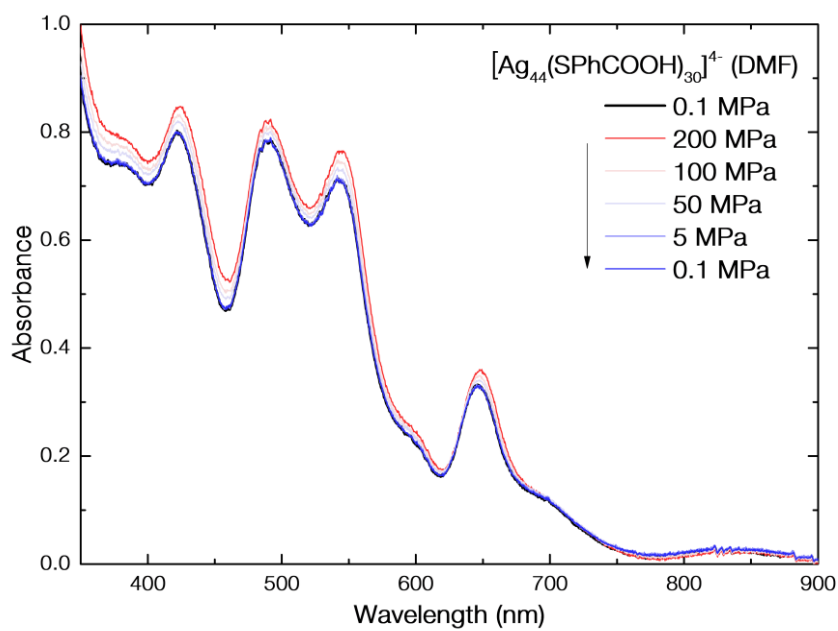

**Figure S4:** UV-Vis absorption spectra of [Ag$_{44}$(SPhCOOH)$_{30}$]$^{4-}$ in DMF upon decreasing pressure from 200 MPa to atmospheric pressure (red to blue). The initial absorption spectra of [Ag$_{44}$(SPhCOOH)$_{30}$]$^{4-}$ is shown in black. This nanocluster species returns to the same level of absorption and spectral energies after increased isotropic pressure, indicating no formation of insoluble aggregates or decomposition of cluster.

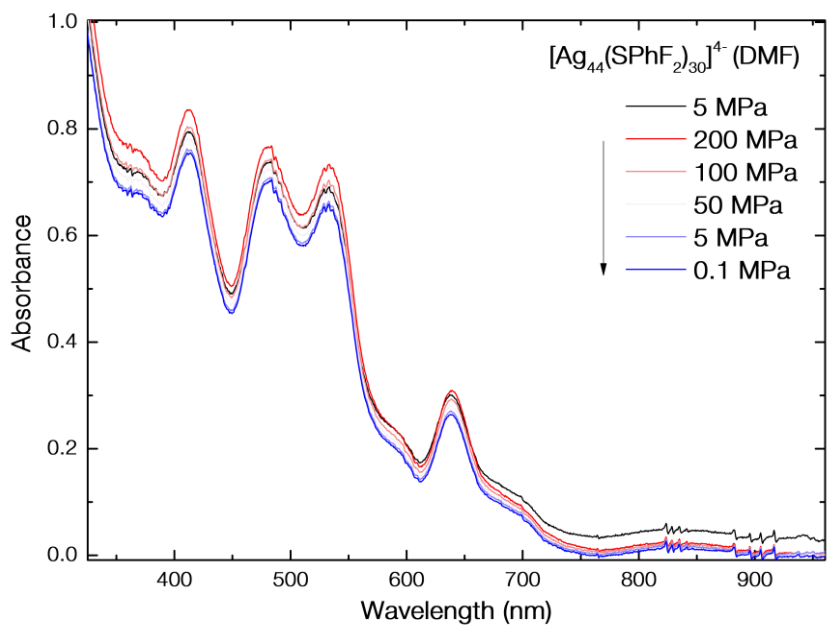

**Figure S5:** UV-Vis absorption spectra of [Ag$_{44}$(SPhF$_2$)$_{30}$]$^{4-}$ in DMF upon decreasing pressure from 200 MPa to atmospheric pressure (red to blue). The initial absorption spectra of [Ag$_{44}$(SPhF$_2$)$_{30}$]$^{4-}$ is shown in black. This nanocluster species exhibits a decrease in absorbance relative to the initial absorption spectrum upon return to atmospheric pressure, indicating that insoluble aggregates may have formed in solution with increased pressure.

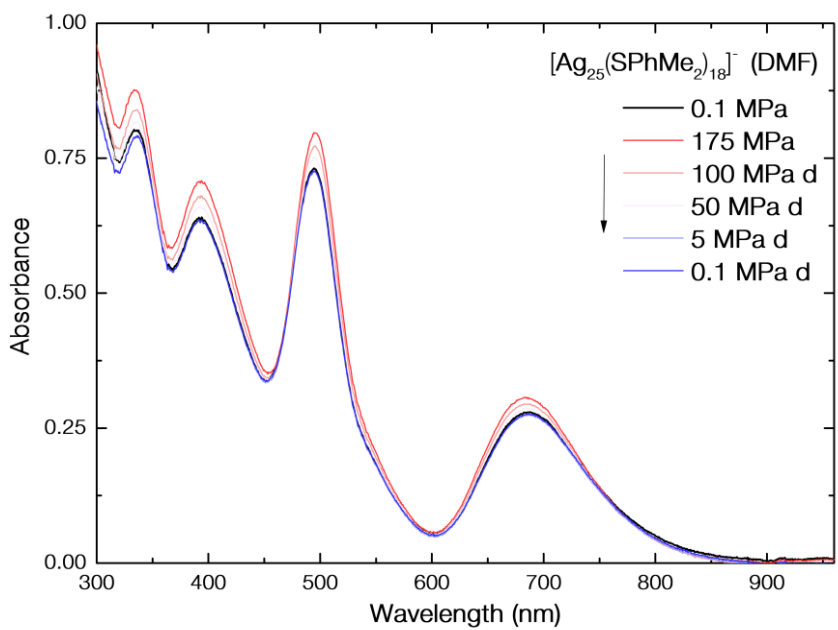

**Figure S6**: UV-Vis absorption spectra of [Ag$_{25}$(SPhMe$_2$)$_{18}$]$^-$ in DMF upon decreasing pressure from 200 MPa to atmospheric pressure (red to blue). The initial absorption spectra of [Ag$_{25}$(SPhMe$_2$)$_{18}$]$^-$ is shown in black. This nanocluster species returns to the same level of absorption and spectral energies after increased isotropic pressure, indicating no formation of insoluble aggregates or decomposition of cluster.

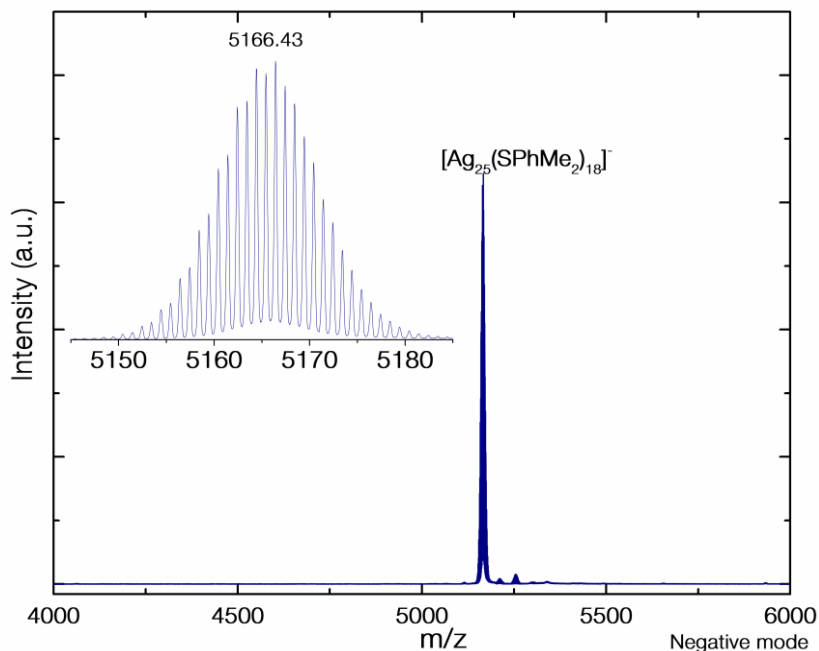

**Figure S7:** Negative mode nanospray MS of [Ag$_{25}$(SPhMe$_2$)$_{18}$]$^-$ nanoclusters synthesised revealing the parent ion peak at m/z 5166 with an interval spacing of ~1 dalton indicating the singularly charged species (inset).

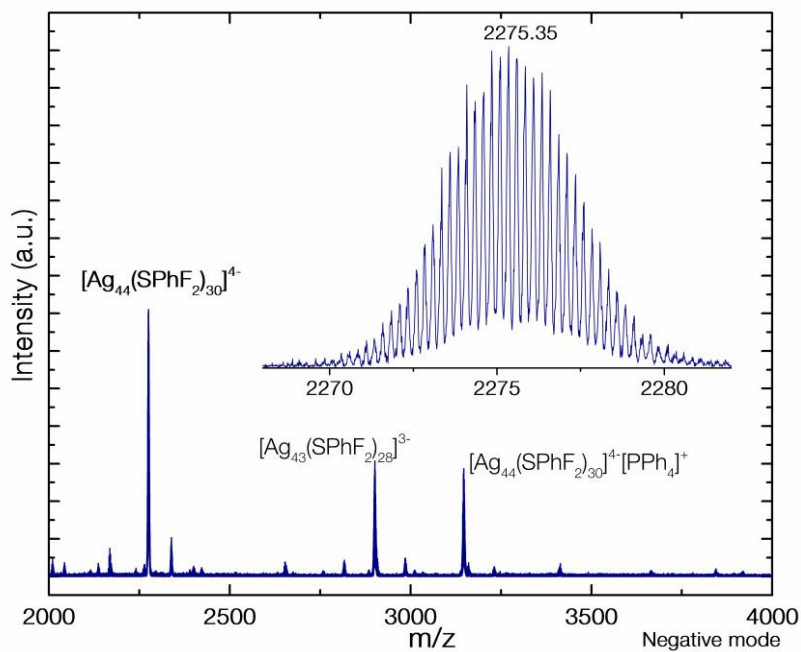

**Figure S8:** Negative mode nanospray MS of [Ag$_{44}$(SPhF$_2$)$_{30}$]$^{4-}$ nanoclusters synthesised revealing the parent ion peak at m/z 2275 with an interval spacing of ~0.25 Dalton indicating the quadruply charged species (inset). Due to the ionisation process of the technique, fragmented species were identified in the spectra such as [Ag$_{43}$(SPhF$_2$)$_{28}$]$^{3-}$ as well adduct species [Ag$_{44}$(SPhF$_2$)$_{30}$]$^{4-}$[PPh$_4$]$^+$. Several unidentifiable peaks are observed in low concentrations.

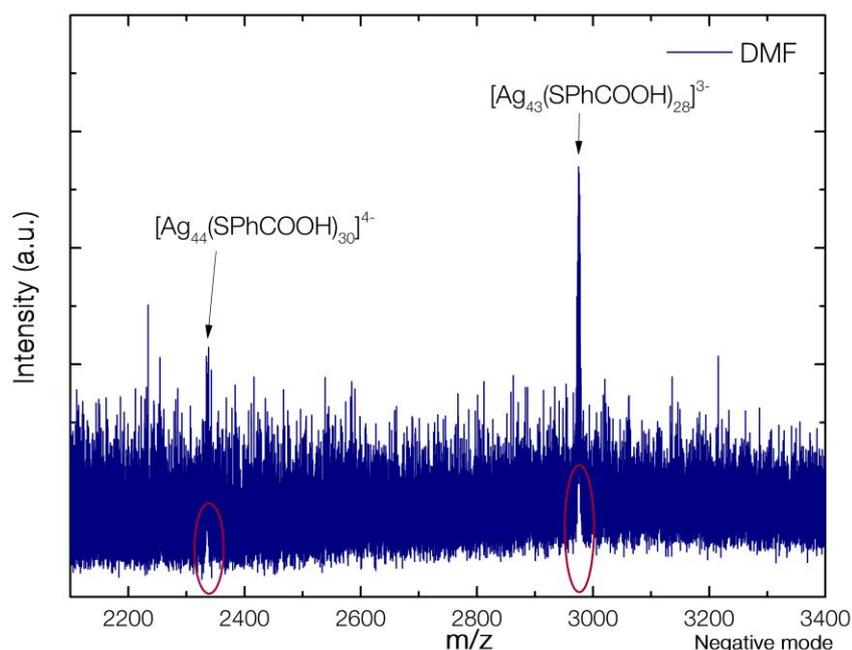

**Figure S9:** Concatenated negative mode nanospray MS of $[Ag_{44}(SPhCOOH)_{30}]^{4-}$ nanoclusters synthesised revealing the parent ion peak at m/z 2336. Nanospray mass spectra revealed no peaks for this nanocluster, only concatenation of $[Ag_{44}(SPhCOOH)_{30}]^{4-}$ nanospray MS data over 5 minutes revealed its presence. After concatenation of data, the true peaks can be discerned by the inflections of the noise baseline highlighted in the figure. A fragmented species is also observed at higher intensity indicating the nanolcuster is more likely to be electrostatically destabilised and fragmented during the MS analysis.

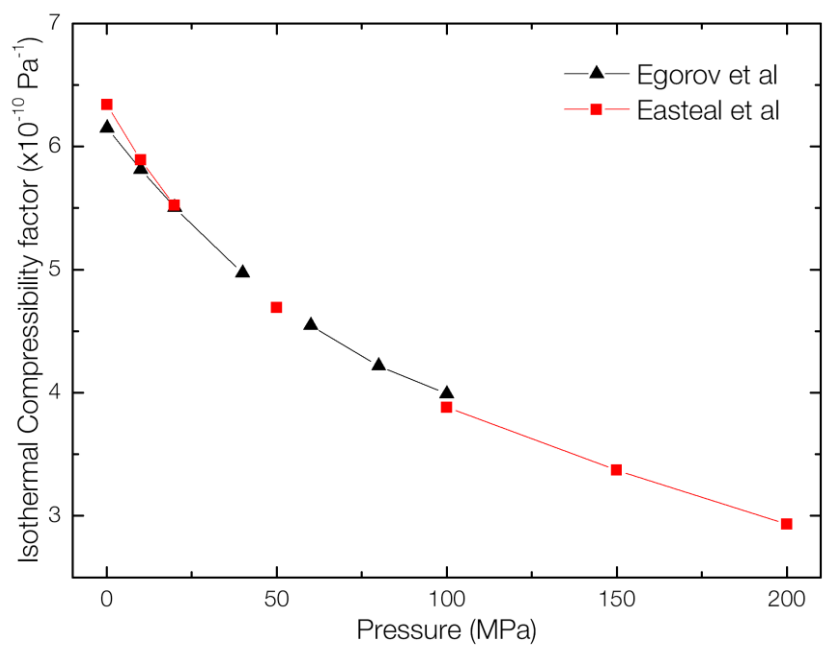

**Figure S9:** Isothermal compressibility of DMF under pressure at 15 °C taken from *Easteal et al* [47] and *Egorov et al* [48]. The change in solvent volume was calculated using the equation *Pressure*(Pa)\**volume* ($m^3$)\**isothermal compressibility factor* ($Pa^{-1}$).